\documentclass[12pt,twoside]{article}

\evensidemargin=\oddsidemargin
\def\Title#1#2#3{%
    \baselineskip=18pt
    \begin{center}
          {\large\bf\uppercase{#1} \\ }
          \bigskip\bigskip
          {#2} \\
          {#3} \\
    \end{center}}
\long\def\Abstract#1{%
         \bigskip
         \parbox{0.93\textwidth}{%
                 \begin{center}
                       {\bf Abstract} \\
                 \end{center}
                 \medskip{\baselineskip=14pt #1}
                 \vss}
         \bigskip}
\newcommand{\myfigure}[1]{\begin{figure*} \centering
        \framebox[150mm]{\epsfxsize=148mm\epsfbox{#1}}
        \par\medskip
        Fig. \thefigure
        \addtocounter{figure}{1}
        \end{figure*}}
\input epsf

\makeatletter
\renewcommand{\section}%
 {\@startsection{section}{1}{0pt}%
  {-3.25ex plus -1ex minus -.2ex}{1.5ex plus .2ex}%
  {\vspace*{5mm}\raggedright\large\bf }}
\renewcommand{\thesection}{\arabic{section}.}
\renewcommand{\thefigure}{\arabic{figure}.}
\@addtoreset{equation}{section}
\renewcommand{\@eqnnum}{(\thesection\theequation)}
\renewcommand{\p@equation}{\thesection}
\makeatother

\begin{document}

\vspace*{1cm}

\Title{CHANGING THE HILBERT SPACE STRUCTURE \\
AS A CONSEQUENCE OF GAUGE TRANSFORMATIONS \\
IN `EXTENDED PHASE SPACE' VERSION \\
OF QUANTUM GEOMETRODYNAMICS}%
{T. P. Shestakova}%
{Department of Theoretical and Computational Physics, Rostov State University, \\
Sorge St. 5, Rostov-on-Don 344090, Russia \\
E-mail: {\tt shestakova@phys.rsu.ru}}

\Abstract{In the earlier works on quantum geometrodynamics in extended
phase space it has been argued that a wave function of the Universe
should satisfy a Schr\"odinger equation. Its form, as well as a
measure in Schr\"odinger scalar product, depends on a gauge
condition (a chosen reference frame). It is known that the geometry
of an appropriate Hilbert space is determined by introducing the
scalar product, so the Hilbert space structure turns out to be in a
large degree depending on a chosen gauge condition. In the present
work we analyse this issue from the viewpoint of the path integral
approach. We consider how the gauge condition changes as a result
of gauge transformations. In this respect, three kinds of gauge
transformations can be singled out: Firstly, there are residual
gauge transformations, which do not change the gauge condition. The
second kind is the transformations whose parameters can be related
by homotopy. Then the change of gauge condition could be described
by smoothly changing function. In particular, in this context time
dependent gauges could be discussed. We also suggest that this kind
of gauge transformations leads to a smooth changing of solutions to
the Schr\"odinger equation. The third kind of the transformations
includes those whose parameters belong to different homotopy
classes. They are of the most interest from the viewpoint of
changing the Hilbert space structure. In this case the gauge
condition and the very form of the Schr\"odinger equation would
change in discrete steps when we pass from a spacetime region with
one gauge condition to another region with another gauge condition.
In conclusion we discuss the relation between quantum gravity and
fundamental problems of ordinary quantum mechanics.}

\section{Introduction}
One of unsolved problems of the Wheeler - DeWitt quantum
geometrodynamics is that of Hilbert space structure. The Wheeler -
DeWitt quantum geometrodynamics was the first attempt of
constructing a quantum theory of the Universe as a whole, however,
if its Hilbert space structure is not rigorously determined, one
cannot consider it as full and consistent, as well as any quantum
theory.

The reasons, why this problem cannot be solved in the framework of
the Wheeler -- DeWitt quantum geometrodynamics, are closely
connected with the fact that it was thought of as a gauge invariant
theory. According to the original idea of Wheeler, a wave function
of the Universe, which is a basic object in quantum
geometrodynamics, must depend on 3-geometry of a manifold
$\cal{M}$. In other words, if $\rm Riem(\cal{M})$ is the space of
all Riemannian metrics on $\cal{M}$, and $\rm Diff(\cal{M})$ is
diffeomorphism group, the wave function must be defined on the
so-called superspace of all 3-geometries, or factor space
$\rm Riem(\cal{M})/\rm Diff(\cal{M})$ \cite{Wheeler,DeWitt1}. One
possible way to express this dependence would be to regard the wave
function as a function of an infinite set of geometrical invariants
\cite{DeWitt2}. It is not clear, however, how to put this idea into
practice. Actually, the wave function depends on a 3-metric, and it
was believed that, if the wave function satisfied a quantum version
of gravitational constraints, it would ensure its dependence on
3-geometry only. The very requirement for the wave function to
satisfy the Wheeler -- DeWitt, but not a Schr\"odinger, equation
leads to the problem of Hilbert space, in particular, it is
questionable how an inner product of state vectors should be
determined (for a recent review on related problems, see
\cite{SS1,SS2}). On the other hand, the Wheeler -- DeWitt quantum
geometrodynamics is based on Arnowitt -- Deser -- Misner (ADM)
formalism, and, as some authors have emphasized \cite{MM1,MM2,MM3},
the latter is equivalent to some kind of gauge fixing, so there is
the inconsistency between appealling to ADM formalism and the
requirement for a wave function to be invariant under
diffeomorphism group transformations.

In this work I shall discuss another approach to quantum
geometrodynamics, namely, quantum geometrodynamics in extended
phase space \cite{SSV1,SSV2,SSV3,SSV4}. The main features of this
approach were presented on the previous PIRT conference
\cite{Shest1}. As was shown in \cite{Shest1}, in the ``extended
phase space'' approach a physical part of the wave function
satisfies a Schr\"odinger equation, whose form, as well as a
measure in Schr\"odinger inner product, depends on a gauge
condition, or a chosen reference frame (the basic formulae will be
repeated in Section 2). The situation can be illustrated by the
following scheme (Fig. 1). All metrics $g_{\mu\nu}$ related by
gauge transformations are unified into an equivalence class
representing dynamics of some 3-geometry. Two metrics $g_{\mu\nu}$
and $g'_{\mu\nu}$, which can be obtained from each other by a
coordinate transformation, correspond to the same geometry, but may
answer to various gauge conditions. In this case in our approach
different gauge conditions correspond to different physical
Hamiltonians, say, $H_1$ and $H_2$. Every of the Hamiltonians acts
in its own Hilbert space with a measure in inner product defined by
a chosen gauge condition. Thus we come to the following question:
How gauge transformations could change the structure of Hilbert
space? To answer it, we shall consider in Section 3 three kinds of
gauge transformations: residual gauge transformations, those whose
parameters related by homotopy and those whose parameters belong to
different homotopy classes. In Section 4 we shall point to some
relation between the problems arising in quantum geometrodynamics
and the problem of reduction of a wave function in ordinary quantum
mechanics, which has been discussed up till now by eminent
physicists (see, for example, \cite{Penrose1,Penrose2,Prig1}).

\myfigure{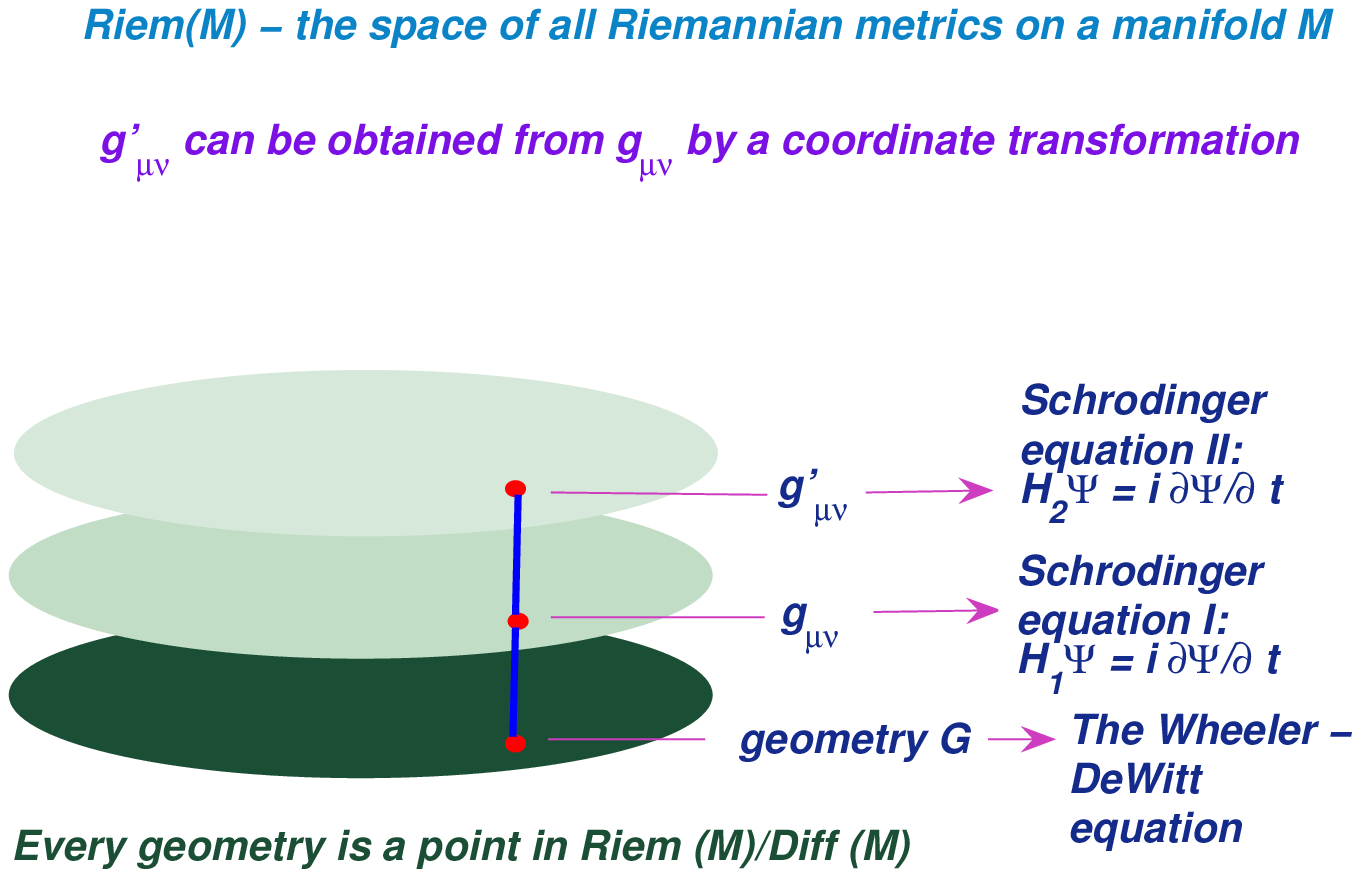}

\section{The Hilbert space in ``extended phase space'' version of quantum
geometrodynamics}
In \cite{Shest1} we considered a simple minisuperspace model with
the action
\begin{equation}
\label{action}
S=\!\int\!dt\,\biggl\{
  \displaystyle\frac12 v(\mu, Q)\gamma_{ab}\dot{Q}^a\dot{Q}^b
  -\frac1{v(\mu, Q)}U(Q)
  +\pi_0\left(\dot\mu-f_{,a}\dot{Q}^a\right)
  -i w(\mu, Q)\dot{\bar\theta}\dot\theta\biggr\}.
\end{equation}
where $Q=\{Q^a\}$ are physical variables, $\theta,\,\bar\theta$ are
the Faddeev -- Popov ghosts and $\mu$ is a gauge variable, its
parameterization being determined by the function $v(\mu,\,Q)$. In
simple cases $\mu$ can be bound to the scale factor $a$ and the lapse
function $N$ by the relation $\displaystyle\frac{a^3}{N}=v(\mu,\,Q)$.
\begin{equation}
\label{w_def}
w(\mu, Q)=\frac{v(\mu, Q)}{v_{,\mu}};\quad
v_{,\mu}\stackrel{def}{=}\frac{\partial v}{\partial\mu}.
\end{equation}
We used a differential form of gauge conditions
\begin{equation}
\label{frame_A}
\mu=f(Q)+k;\quad
k={\rm const},
\end{equation}
namely,
\begin{equation}
\label{diff_form}
\dot{\mu}=f_{,a}\dot{Q}^a,\quad
f_{,a}\stackrel{def}{=}\frac{\partial f}{\partial Q^a}.
\end{equation}
The wave function is defined on extended configurational space with
the coordinates $\mu,\,Q,\,\theta,\,\bar\theta$. In ``extended
phase space'' version of quantum geometrodynamics we quantize ghost
and gauge gravitational degrees of freedom on an equal basis with
physical degrees of freedom. The motivation for it was that it is
impossible to separate gauge, or ``non-physical'' degrees of
freedom from physical ones if the system under consideration does
not possess asymptotic states, and it is indeed the case for a
closed universe as well as in a general case of nontrivial
topology. Then, we come to the Schr\"odinger equation, which is
derived from a path integral with the effective action
(\ref{action}) without asymptotic boundary conditions by the
standard well-definite Feynman procedure, and which is {\it a
direct mathematical consequence} of the path integral.

\begin{equation}
\label{SE1}
i\,\frac{\partial\Psi(\mu,Q,\theta,\bar\theta;\,t)}{\partial t}
 =H\Psi(\mu,\,Q,\,\theta,\,\bar\theta;\,t),
\end{equation}
where
\begin{equation}
\label{H}
H=-\frac i w\frac{\partial}{\partial\theta}
   \frac{\partial}{\partial\bar\theta}
  -\frac1{2M}\frac{\partial}{\partial Q^{\alpha}}MG^{\alpha\beta}
   \frac{\partial}{\partial Q^{\beta}}
  +\frac1v(U-V);
\end{equation}
\begin{equation}
\label{M}
M(\mu, Q)=v^{\frac K2}(\mu, Q)w^{-1}(\mu, Q);
\end{equation}
\begin{equation}
\label{Galpha_beta}
G^{\alpha\beta}=\frac1{v(\mu, Q)}\left(
 \begin{array}{cc}
  f_{,a}f^{,a}&f^{,a}\\
  f^{,a}&\gamma^{ab}
 \end{array}
 \right);\quad
\alpha,\beta=(0,a);\quad
Q^0=\mu,
\end{equation}
$M$ is the measure in inner product, $K$ is a number of physical
degrees of freedom, $V$ is a quantum correction to the potential
$U$, that depends on the chosen parameterization and gauge
\cite{Shest1}. The Schr\"odinger equation (\ref{SE1}) gives {\it a
gauge-dependent description of the Universe}. The general solution
to the equation (\ref{SE1}) is
\begin{equation}
\label{GS-A}
\Psi(\mu,\,Q,\,\theta,\,\bar\theta;\,t)
 =\int\Psi_k(Q,\,t)\,\delta(\mu-f(Q)-k)\,(\bar\theta+i\theta)\,dk.
\end{equation}
It can be interpreted in the spirit of Everett's ``relative state''
formulation: Each element of the superposition (\ref{GS-A})
describe a state in which the only gauge degree of freedom $\mu$ is
definite, so that time scale is determined by processes in the
physical subsystem through functions $v(\mu,\,Q)$, $f(Q)$ while the
function $\Psi_k(Q,\,t)$ describes a state of the physical
subsystem for a reference frame fixed by the condition
(\ref{frame_A}). It is a solution to the Schr\"odinger equation with
a gauge-dependent physical Hamiltonian $H_{(phys)}$:
\begin{equation}
\label{phys.SE}
i\,\frac{\partial\Psi_k(Q;\,t)}{\partial t}
 =H_{(phys)}\Psi_k(Q;\,t),
\end{equation}
\begin{equation}
\label{phys.H-A}
H_{(phys)}=\left.\left[-\frac1{2M}\frac{\partial}{\partial Q^a}
  \frac1v M\gamma^{ab}\frac{\partial}{\partial Q^b}
 +\frac1v (U-V)\right]\right|_{\mu=f(Q)+k}.
\end{equation}
Solutions to Eq.(\ref{phys.SE}) make a basis in the Hilbert space
of states of the physical subsystem:
\begin{equation}
\label{stat.states}
H_{(phys)}\Psi_{kn}(Q)=E_n\Psi_{kn}(Q);
\end{equation}
\begin{equation}
\label{stat.WF}
\Psi_k(Q,\,t)=\sum_n c_n\Psi_{kn}(Q)\exp(-iE_n t).
\end{equation}
As one can see, the spectrum and eigenfunctions of the operator
$H_{(phys)}$ will depend on a chosen gauge condition. The
dependence of the measure in the physical subspace on this gauge
results from the normalization condition for the wave function
(\ref{GS-A}):
\begin{eqnarray}
&&\hspace{-1cm}
\int\Psi^*(\mu,\,Q,\,\theta,\,\bar\theta;\,t)\,
 \Psi(\mu,\,Q,\,\theta,\,\bar\theta;\,t)\,M(\mu,\,Q)\,
 d\mu\,d\theta\,d\bar\theta\,\prod_adQ^a=\nonumber\\
&=&\int\Psi^*_k(Q,\,t)\,\Psi_{k'}(Q,\,t)\,
 \delta(\mu-f(Q)-k)\,\delta(\mu-f(Q)-k')\,M(\mu,\,Q)\,
 dk\,dk'\,d\mu\,\prod_adQ^a=\nonumber\\
\label{Psi_norm}
&=&\int\Psi^*_k(Q,\,t)\,\Psi_k(Q,\,t)\,
 M(f(Q)+k,\,Q)\,dk\,\prod_adQ^a=1.
\end{eqnarray}
Therefore, the whole structure of the physical Hilbert space is
formed in a large degree by the chosen gauge condition (reference
frame). {\it One cannot give a consistent quantum description of the
Universe without fixing a certain reference frame, as well as one
cannot find a solution to classical Einstein equations without
imposing some gauge conditions}. The attempt to give a gauge
invariant description of the Universe in the limits of the Wheeler
-- DeWitt quantum geometrodynamics was not successful, and the
problem of Hilbert space is just the fact indicating that this
theory has to be modified.

On the other hand, in the ``extended phase space'' approach we face
another problem, that for every gauge condition we have its own
Hilbert space. Is there any relation between state vectors in these
Hilbert spaces, or between solutions to Schr\"odinger equations
corresponding to various reference frames? How does the structure
of Hilbert space change if one varies a gauge condition? We shall
try to discuss these issues in the next sections.

\section{Path integral and three kinds of gauge transformations}
Let us consider a spacetime manifold ${\cal M}$, which consists of
several regions ${\cal R}_1,\,{\cal R}_2,\,{\cal R}_3,\,\ldots$, in
each of them various gauge conditions $C_1,\,C_2,\,C_3,\,\ldots$
being imposed. It is naturally to think that such regions exist in
a universe with a non-trivial topology. Just for simplicity, one
can assume that boundaries ${\cal S}_1,\,{\cal S}_2,\,\ldots$
between the regions are spacelike and can be labeled by some time
variables $t_1,\,t_2,\,\ldots$ (Fig. 2).

\myfigure{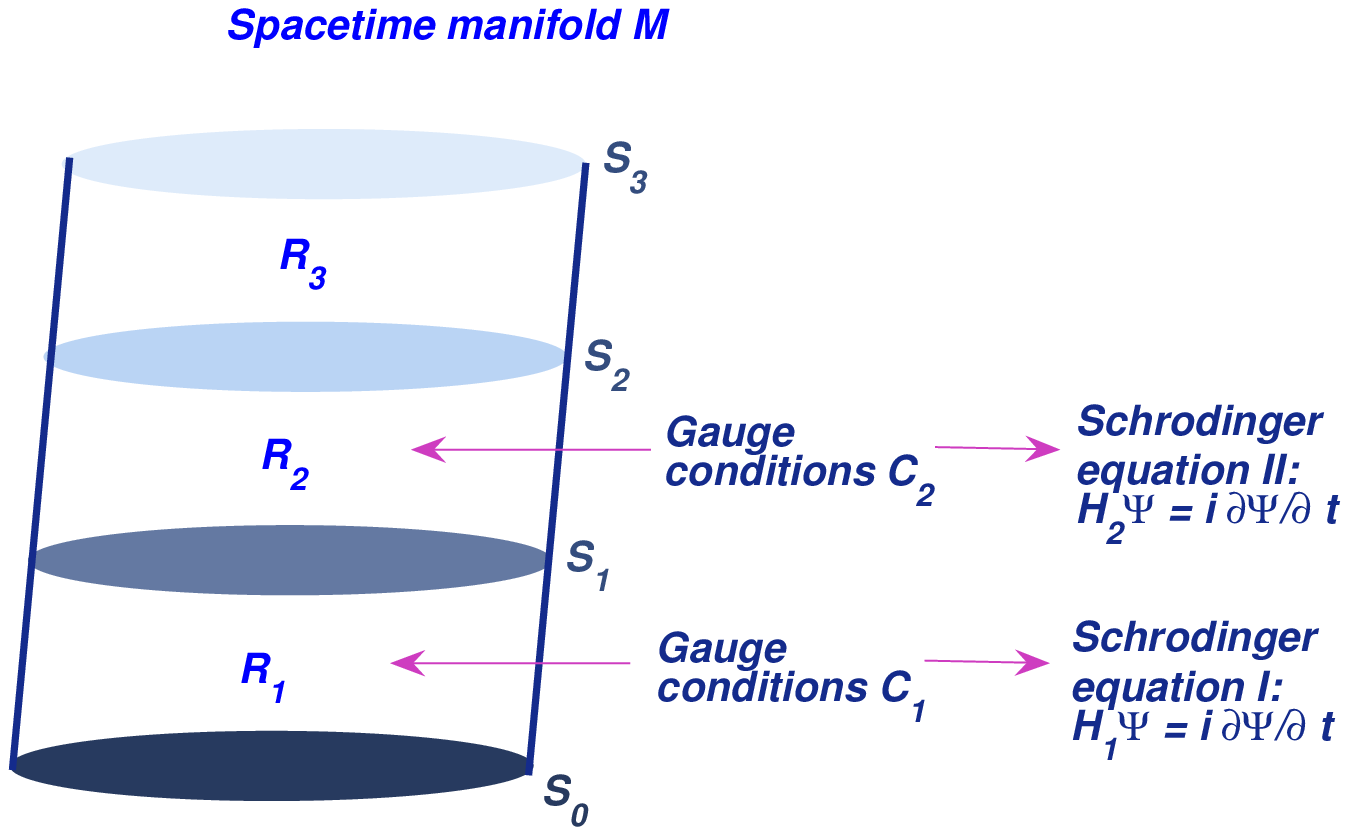}

We would emphasize that the path integral approach allows us to
examine this situation without any generalization of the
formalism. The path integral over the manifold ${\cal M}$ is
\begin{eqnarray}
&&\hspace{-1cm}
\int\exp\left(iS\left[g_{\mu\nu}\right]\right)
 \prod\limits_{x\in{\cal M}}M\left[g_{\mu\nu}\right]
 \prod\limits_{\mu,\,\nu}dg_{\mu\nu}(x)=\nonumber\\
&=&\int\exp\left(iS_{(eff)}
     \left[g_{\mu\nu},\,C_1,\,{\cal R}_1\right]\right)
    \prod\limits_{x\in{\cal R}_1}
     M\left[g_{\mu\nu},\,{\cal R}_1\right]
    \prod\limits_{\mu,\,\nu}dg_{\mu\nu}(x)\times\nonumber\\
&&\times\exp\left(iS_{(eff)}
     \left[g_{\mu\nu},\,C_2,\,{\cal R}_2\right]\right)
    \prod\limits_{x\in{\cal R}_2}
     M\left[g_{\mu\nu},\,{\cal R}_2\right]
    \prod\limits_{\mu,\,\nu}dg_{\mu\nu}(x)\times\nonumber\\
\label{PI1}
&&\hspace{48mm}\times\prod\limits_{x\in{\cal S}_1}
     M\left[g_{\mu\nu},\,{\cal S}_1\right]
    \prod\limits_{\mu,\,\nu}dg_{\mu\nu}(x)\times\ldots
\end{eqnarray}
Here $S_{(eff)}\left[g_{\mu\nu},\,C_1,\,{\cal R}_1\right]$ is the
effective action in the region ${\cal R}_1$ with gauge conditions
$C_1$, which includes gauge fixing and ghosts terms, etc.

From the viewpoint of gauge invariant approach, the path integral
is not to depend on gauge conditions, in other words, we could
write
\begin{eqnarray}
&&\hspace{-1cm}
\int\exp\left(iS\left[g_{\mu\nu}\right]\right)
 \prod\limits_{x\in{\cal M}}M\left[g_{\mu\nu}\right]
 \prod\limits_{\mu,\,\nu}dg_{\mu\nu}(x)=\nonumber\\
\label{PI2}
&=&\int\langle g_{\mu\nu}^{(0)},\,{\cal S}_0
     |g_{\mu\nu}^{(1)},\,{\cal S}_1\rangle
    \langle g_{\mu\nu}^{(1)},\,{\cal S}_1
     |g_{\mu\nu}^{(2)},\,{\cal S}_2\rangle
    \prod\limits_{x\in{\cal S}_1}
     M\left[g_{\mu\nu},\,{\cal S}_1\right]
    \prod\limits_{\mu,\,\nu}dg_{\mu\nu}(x)\times\ldots
\end{eqnarray}
In this case the initial state
$|g_{\mu\nu}^{(0)},\,{\cal S}_0\rangle$, as well as intermediate
states $|g_{\mu\nu}^{(1)},\,{\cal S}_1\rangle$,
$|g_{\mu\nu}^{(2)},\,{\cal S}_2\rangle$, etc. are supposed to be
gauge invariant (i.e. independent on ghosts and gauge conditions).
This assumption would be justified only if all the states were
asymptotic, but it cannot be true at least for the intermediate
states. Moreover, the path integral (\ref{PI2}) needs to be
regularized, that implies imposing gauge condition on the surface
${\cal S}_1$ (\cite{SSV3}; see also \cite{Shest2}). Eq.(\ref{PI2})
is a generalization of the well-known quantum mechanical operation
when one inserts ``a full set of states'' at some moment $t_1$. But
in the present consideration we should bear in mind that the states
in the regions ${\cal R}_1$ and ${\cal R}_2$ belong to different
Hilbert spaces.

Within the region ${\cal R}_1$ the evolution of the physical
subsystem is determined by a unitary operator
$\exp\left[-iH_{1(phys)}\left(t_1-t_0\right)\right]$, where
$H_{1(phys)}$ is a physical Hamiltonian in the region ${\cal R}_1$
with gauge conditions $C_1$. Let at initial time $t_0$ on the
surface ${\cal S}_0$ the state of the system is given by a vector
$|g_{\mu\nu}^{(0)},\,{\cal S}_0\rangle$. Then the state on the
boundary ${\cal S}_1$ reads
\begin{equation}
\label{g1}
|g_{\mu\nu}^{(1)},\,{\cal S}_1\rangle=
 \exp\left[-iH_{1(phys)}\left(t_1-t_0\right)\right]
 |g_{\mu\nu}^{(0)},\,{\cal S}_0\rangle.
\end{equation}
However, if we gone from the region ${\cal R}_1$ to ${\cal R}_2$,
we would find ourselves in another Hilbert space with a basis
formed from eigenfunctions of the operator $H_{2(phys)}$. The
transition to a new basis is not a unitary operation, as follows
from the fact that a measure in the physical subspace depends on
gauge conditions \cite{Shest1,Shest3} (in our minisuperspace model
it is demonstrated by (\ref{Psi_norm})). Denote the operation of
the transition to a new basis in the region ${\cal R}_2$ as
${\cal P}\left({\cal S}_1,\,t_1\right)$. Then the initial state in
the region ${\cal R}_2$ is
\begin{equation}
\label{projection}
{\cal P}\left({\cal S}_1,\,t_1\right)
 \exp\left[-iH_{1(phys)}\left(t_1-t_0\right)\right]
 |g_{\mu\nu}^{(0)},\,{\cal S}_0\rangle
\end{equation}
and
\begin{eqnarray}
|g_{\mu\nu}^{(3)},\,{\cal S}_3\rangle
&=&\exp\left[-iH_{3(phys)}\left(t_3-t_2\right)\right]
 {\cal P}\left({\cal S}_2,\,t_2\right)
 \exp\left[-iH_{2(phys)}\left(t_2-t_1\right)\right]\times\nonumber\\
\label{g3}
&\times&{\cal P}\left({\cal S}_1,\,t_1\right)
 \exp\left[-iH_{1(phys)}\left(t_1-t_0\right)\right]
 |g_{\mu\nu}^{(0)},\,{\cal S}_0\rangle.
\end{eqnarray}

So, {\it at any border ${\cal S}_i$ between regions with different
gauge conditions unitary evolution is broken down. The operators
${\cal P}\left({\cal S}_i,\,t_i\right)$ play the role of projection
operators, which project states obtained by unitary evolution in a
region ${\cal R}_i$ on a basis in Hilbert space in a neighbour
region ${\cal R}_{i+1}$}.

We now turn to different types of gauge transformations. It is
conventionally believed that gauge conditions
\begin{equation}
\label{gen.GC}
F^{\mu}\left[g^{\lambda\rho}(x),\,\theta^{\nu}(x)\right]=0
\end{equation}
should be chosen to fix completely gauge transformation parameters.
Meanwhile, one knows that, in general, these conditions fix gauge
parameters up to residual transformations satisfying the equations
which are consequence of (\ref{gen.GC}):
\begin{equation}
\label{residual}
\delta F^{\mu}\left[g^{\lambda\rho}(x),\,\theta^{\nu}(x)\right]=0
\quad\Rightarrow\quad
A^{\mu}_{\nu}\theta^{\nu}(x)=
 \frac{\delta F^{\mu}}{\delta g^{\lambda\rho}}
 \frac{\delta
 g^{\lambda\rho}}{\delta\theta^{\nu}}\theta^{\nu}(x)=0.
\end{equation}
However, we should not worry about this kind of transformations
since they do not change the conditions (\ref{gen.GC}) and not
affect the structure of Hilbert space.

More interesting are the transformations whose parameters can be
related by homotopy. Consider two gauge conditions
\begin{equation}
\label{gc1}
F^{\mu}_1\left[g^{\lambda\rho}(x),\,\theta^{\nu}_1(x)\right]=0;
\end{equation}
\begin{equation}
\label{gc2}
F^{\mu}_2\left[g^{\lambda\rho}(x),\,\theta^{\nu}_2(x)\right]=0,
\end{equation}
fixing points on a gauge orbit in which a group element is
parameterized by $\theta^{\nu}_1(x)$ and $\theta^{\nu}_2(x)$,
correspondingly. Let us assume that there exists continuous
functions $L^{\nu}(r,\,x)$, so that
\begin{equation}
\label{homotopy1}
L^{\nu}(r,\,x):\quad
L^{\nu}(0,\,x)=\theta^{\nu}_1(x),\quad
L^{\nu}(1,\,x)=\theta^{\nu}_2(x),
\end{equation}
or, more generally,
\begin{equation}
\label{homotopy2}
L^{\nu}(r,\,x):\quad
L^{\nu}(r_1,\,x)=\theta^{\nu}_1(x),\quad
L^{\nu}(r_2,\,x)=\theta^{\nu}_2(x).
\end{equation}
One would say that $\theta^{\nu}_1(x)$ and $\theta^{\nu}_2(x)$
belong to the same homotopy class. Further, we could introduce a
set of gauge conditions
\begin{equation}
\label{GC_set}
F^{\mu}\left[g^{\lambda\rho}(x),\,\theta^{\nu}_r(x);\,r\right]=0:\quad
\theta^{\nu}_r(x)=L^{\nu}(r,\,x),
\end{equation}
and identify $r$ with a time variable $t$. Then, time-dependent
conditions (\ref{GC_set}) could be interpreted as describing a
smooth transition from the gauge (\ref{gc1}) to (\ref{gc2}). Our
ability to impose the set of conditions (\ref{GC_set}) depends on
the structure of group and related to the possibility of
introducing some special coordinates in group space \cite{DeWitt1}.
In our simple minisuperspace model before gauge fixing the action
is invariant under one-parametric Abelian group of transformations
\begin{equation}
\label{Ab.group}
\delta t=\theta(t);\quad
\delta\mu=w(\mu,\,Q)\dot\theta-\dot\mu\theta;\quad
\delta Q^a=-\dot Q^a\theta,
\end{equation}
so that any time-dependent gauge condition
\begin{equation}
\label{time_gauge}
\mu=f(Q,\,t)+k;\quad
k={\rm const},
\end{equation}
would satisfy the above assumption.

Any canonical time-dependent gauge constrained physical variables
and their momenta
\begin{equation}
\label{chi_gauge}
\chi\left(Q,\,P,\,t\right)=0
\end{equation}
can be reduced by Dirac-like procedure to the form similar to
(\ref{time_gauge}). In the canonical approach, choosing a simple
parameterization $v(\mu,\,Q)=\displaystyle\frac1{\mu}$, one would
find that the canonical Hamiltonian of the system $H$ is proportional
to the secondary constraint $T$:
\begin{equation}
\label{can.Hamilt.}
H=\mu T=\mu\left[\frac12 P_a P^a+U(Q)\right].
\end{equation}
From the requirement of the conservation of (\ref{chi_gauge}) in
time \cite{AB} one obtains
\begin{equation}
\label{chi_dot}
\frac{d\chi}{d t}=
 \frac{\partial\chi}{\partial t}+\mu\{\chi,\,T\}=0;
\end{equation}
\begin{equation}
\label{mu_gauge}
\mu=-\frac{\partial\chi}{\partial t}\{\chi,\,T\}^{-1}=
 \tilde f\left(Q,\,P,\,t\right),
\end{equation}
the letter can be presented in a differential form. We would like
to emphasize here that, though quantization schemes using canonical
time-dependent gauges (\ref{chi_gauge}) are believed to be
equivalent to gauge invariant Dirac quantization \cite{AB}, from
the viewpoint of our approach imposing such gauge conditions
implies gauge-dependent structure of physical Hilbert space.

The formalism developed in \cite{SSV1,SSV2,SSV3,SSV4} can be
generalized to gauge conditions explicitly depending on time. The
pass integral approach includes some skeletonization procedure,
which implies approximation of the gauge on each time interval
$\left[t_i,\,t_{i+1}\right]$. In the simplest situation, we could
assume that the change of gauge condition in each time interval is
given by a function
\begin{equation}
\label{delta_f}
\delta f_i(Q)=\alpha f_i(Q),
\end{equation}
$\alpha$ is a small parameter, so that the gauge condition is a
step-like function
\begin{equation}
\label{step.func}
\mu=f(Q)+\sum\limits_i\alpha f_i(Q)\theta\left(t-t_i\right)+k
\end{equation}
in the sense that in each interval $\left[t_n,\,t_{n+1}\right]$ the
gauge condition does not depend on time:
\begin{equation}
\label{time_int}
\left[t_n,\,t_{n+1}\right]:\quad
\mu=f(Q)+\sum\limits_{i=0}^{n-1}\alpha f_i(Q)+\delta f_n(Q)+k.
\end{equation}
Thus, we have come to the case of a small variation of gauge
condition that was discussed in \cite{Shest3}. As was shown in
\cite{Shest3}, this small variation gives rise to additional terms
in a physical Hamiltonian, these terms being non-Hermitian in
respect to original physical subspace before variation. In our
time-dependent case it means that at every moment of time we have a
Hamiltonian, which acts in its own ``instantaneous'' Hilbert space.
The instantaneous Hamiltonian is a Hermitian operator at each
moment, but one should think of it as a non-Hermitian operator in
respect to a Hilbert space one had at a previous moment. The
situation is different from what we have in ordinary quantum
mechanics for a time-dependent Hamiltonian that acts at every
moment in the same Hilbert space whose measure does not change in
time. An analogy can be drawn between our situation and particle
creation in nonstationary gravitational field when we also have an
instantaneous Hamiltonian and instantaneous Fock basis \cite{Grib}.

Smooth changing of a gauge condition in time implies that solutions
to the Schr\"odinger equation for physical part of wave function
also change in a continuous and smooth manner. Another situation we
face when gauge conditions in two regions fix gauge parameters
which belong to different homotopy classes, and, as a rule,
spacetime coordinates in these regions being related by a singular
transformation. Then the gauge condition and the very form of the
Schr\"odinger equation change in discrete steps when one passes
from a spacetime region with some gauge condition to a region with
another gauge condition. This case is of the most interest from the
viewpoint of changing the Hilbert space structure and the most
difficult to treat. In any case, an initial state in a region
${\cal R}_i$, resulting from its preceding evolution, should be
written as a superposition of states in a new Hilbert space in
${\cal R}_i$. There arise a number of questions, like: Will this
superposition of states be stable? Could the breakdown of unitarity
give rise to some kind of irreversibility? Could we define the
change of entropy of the physical system when going to a region
with different gauge conditions? Possible answers depend on a
chosen model and require new non-perturbation methods.

\section{Conclusion: the problem of wave function reduction and Quantum
Gravity}
As was pointed out by von Neumann \cite{Neumann}, in quantum
mechanics one deals with two different processes, namely, unitary
evolution of a physical system in time described by the
Schr\"odinger equation, and reduction of wave function of the
physical system under observation. The whole evolution of the
system can be presented by the formula
\begin{eqnarray}
|\Psi\left(t_N\right)\rangle
&=&U\left(t_N,\,t_{N-1}\right){\cal P}\left(t_{N-1}\right)
   U\left(t_{N-1},\,t_{N-2}\right)\ldots\times\nonumber\\
\label{evolution}
&\times&\ldots U\left(t_3,\,t_2\right){\cal P}\left(t_2\right)
   U\left(t_2,\,t_1\right){\cal P}\left(t_1\right)
   U\left(t_1,\,t_0\right)|\Psi\left(t_0\right)\rangle,
\end{eqnarray}
where ${\cal P}\left(t_i\right)$ are projection operators
corresponding to observation at moments $t_1$, $t_2$,
$t_3,\,\ldots$, $t_{N-1}$ (see, for example, \cite{Mensky}). There
arises an analogy between the formulae (\ref{g3}) and
(\ref{evolution}): Indeed, we interpret any reference frame as a
measuring instrument representing the observer in quantum
geometrodynamics. Gauge conditions define interaction between the
measuring instrument (reference frame) and the physical subsystem
of the Universe. Changing the interaction with the measuring
instrument makes us go to another basis in a Hilbert space and,
even more, to another Hilbert space.

It enables us to hope to throw a new look to the central quantum
mechanical problem of wave function reduction. Roger Penrose
pointed out time and again that a solution of this problem, as well
as understanding of irreversibility of physical processes, must be
closely related with the progress in constructing quantum theory of
gravity. In his books \cite{Penrose1,Penrose2} Penrose proposed a
mechanism anticipating a choice among spacetime geometries, each of
them corresponding to an element of quantum superposition. Details
of the mechanism had not been elaborated enough, and this proposal
was strongly criticized by Hawking \cite{Hawking}. However, the
main idea that quantum gravity may help in deeper understanding of
quantum mechanics seems to be fruitful. In our ``extended phase
space'' approach we face the situation when the breakdown of
unitary evolution of a physical system naturally follows from the
very structure of the theory -- we do not need to introduce ``by
hands'' some special interaction, which would result in the
breakdown of unitarity. In its turn, it is connected to the
irreversibility of measuring processes. According to the opinion of
another famous scientist, Ilya Prigogine, symmetric in time quantum
dynamics described by the Schr\"odinger equation should be
generalized to involve irreversible processes. To do it, one has to
extend the class of admissible quantum operators beyond Hermitian
operators and include non-unitary transformations of state vectors
or density matrices (\cite{Prig1}; see also his Nobel prize lecture
\cite{Prig2}). On the other side, in quantum mechanics one could
examine models of interaction with a measuring instrument in which
coordinates of a physical system are bound to coordinates of the
instrument by means of some constraints, the latter ones are, in a
sense, ``gauge conditions'' like those we have considered in our
model with finite number degrees of freedom. Similar models of
interaction with the instrument had been explored yet by von
Neumann \cite{Neumann}. In future, some general points in quantum
mechanical and quantum gravitational models of interaction may be
revealed.

It may seem that there are more question than answers in this
report. However, we have discussing physical interpretations of
Relativity Theory already for a hundred years. So it is not
surprisingly that attempts of its unification with quantum theory
pose even more fundamental and intriguing questions, which still
have been waiting for their resolution.

\end{document}